\documentstyle[worldsci,epsf,fleqn]{article}

\newcommand{\be}[1]{\begin{equation} \label{(#1)}}
\newcommand{\ee}{\end{equation}}
\newcommand{\ba}[1]{\begin{eqnarray} \label{(#1)}}
\newcommand{\ea}{\end{eqnarray}}

\begin{document}

\title{\Large{\bf A general parametrization for the long-range part of neutrinoless 
double beta decay
\footnote[2]{Talk presented by H. P\"as at the 
Erice School on Nuclear Physics, 19th course "Neutrinos in Astro, Particle and 
Nuclear Physics", Erice, Italy, 16-24 September 1997}
}}
\bigskip
\author{H. P\"as
\footnote[1]{present address: Laboratori Nazionali del Gran Sasso
(INFN), 67010 Assergi (AQ), Italy;\\
E-mail:
Heinrich.Paes@mpi-hd.mpg.de}, M. Hirsch, 
S.G. Kovalenko
\footnote[3]{On leave from Joint Institute for Nuclear Research, 
Dubna, Russia}, 
H.V. Klapdor--Kleingrothaus\\
\medskip
{\it Max--Planck--Institut f\"ur Kernphysik,\\ P.O. Box 103980,
D--69029 Heidelberg, Germany}}
%
\maketitle
\bigskip

\begin{abstract}
\noindent
Double beta decay has been proven to be a powerful tool to 
constrain
$B-L$ violating physics beyond the standard model. 
We present a representation
for the long--range part of the general $0\nu\beta\beta$ decay rate
allowed by Lorentz--invariance. Combined with the short range part
this general parametrization in terms of effective $B-L$ violating 
couplings will provide the $0\nu\beta\beta$ limits on arbitrary
lepton number violating theories.
\end{abstract} 
\bigskip

\section{General Formalism}
We consider the long--range part of neutrinoless double beta decay with 
two vertices, which are pointlike at the Fermi scale, and exchange of a
light neutrino in between. The general Lagrangian can be written in terms of
effective couplings $\epsilon^{\alpha}_{\beta}$, which correspond to the
pointlike vertices at the Fermi scale so that Fierz rearrangement is 
applicable:
\be{1}
{\cal L}=\frac{G_F}{\sqrt{2}}\{
j_{V-A}^{\mu}J^{\dagger}_{V-A,\mu}+ \sum_{\alpha,\beta} 
 ^{'}\epsilon_{\alpha}^{\beta}j_{\beta} J^{\dagger}_{\alpha}\}
\ee
with the combinations of hadronic and leptonic Lorentz currents of defined
helicity 
$\alpha,\beta=V-A,V+A,S+P,S-P,T_L=2 T P_L,T_R=2 T P_R$ and the usual 
left-- and right
handed projectors $P_{L/R}=\frac{1 \mp \gamma_5}{2}$. 
The sum runs over all contractions allowed by Lorentz--invariance,
except for $\alpha=\beta=V-A$. 
Evaluating ``on axis'' one assumes only one of the 
$\epsilon_{\alpha}^{\beta}$
unequal zero and arrives at a general double beta decay amplitude proportional
to
\be{2}
T({\cal L}_{(1)} {\cal L}_{(2)})=\frac{G_F^2}{2}
T\{j_{V-A}J^{\dagger}_{V-A}j_{V-A}J^{\dagger}_{V-A}
  + \epsilon_{\alpha}^{\beta}j_{\beta} J^{\dagger}_{\alpha}j_{V-A}
J^{\dagger}_{V-A} 
+ (\epsilon_{\alpha}^{\beta})^2 j_{\beta} J^{\dagger}_{\alpha} j_{\beta} 
J^{\dagger}_{\alpha}\}.
\ee
While the first term corresponds to the standard model (SM)
like neutrino exchange
and the 3rd term $\sim (\epsilon_{\alpha}^{\beta})^2$ can be neglected, only
the 2nd term is phenomenological interesting. For this term one has to
consider two general cases:

1) The leptonic SM $V-A$ current meets a left--handed non SM current 
   $j_{\beta}$ with $\beta=S-P,T_L$. This contribution is proportional
   to the unknown neutrino Majorana mass $m_{\nu}\leq \sim 
   0.5 eV$, for which 
   no lower bound exists. Therefore no limits on the parameters 
   $\epsilon_{\alpha}^{\beta}$ can be derived.

2) The leptonic SM $V-A$ current meets a right--handed non SM current
   $j_{\beta}$ with $\beta=S+P,V+A,T_R$. This contribution is proportional
   to the neutrino momentum $q\simeq p_F \simeq 100 MeV$ with the nuclear 
   Fermi momentum $p_F$, and thus will produce stringent limits on
   $\epsilon_{\alpha}^{\beta}$.

Taking these considerations into account, we are left with three interesting 
contributions discussed in the following sections. 
With the present half-life limit of the Heidelberg--Moscow 
experiment $T_{1/2}^{0\nu\beta\beta}>1.1 \cdot 10^{25} y$ \cite{HM97} 
and
\be{t12}
[T_{1/2}^{0\nu\beta\beta}]^{-1}=(\epsilon_{\alpha}^{\beta})^2 G_{0k} |ME|^2
\ee
where $G_{0k}$ denotes the phase space factors defined in \cite{Doi85} 
and $|ME|$ the nuclear 
matrix elements, limits on the $\epsilon_{\alpha}^{\beta}$ can be 
derived.

\section{Calculational details and preliminary limits}
\subsection{SM meets $j_{V+A}J^{\dagger}_{V+A}$ and $j_{V+A}
J^{\dagger}_{V-A}$}
This contribution has been considered already in the context of left--right 
symmetric models \cite{Doi85,Hir96c}. 
For sake of completeness we repeat the results here in our notation:
$\epsilon^{V+A}_{V+A}< 8.2 \cdot 10^{-7}$, 
$\epsilon^{V+A}_{V-A}< 4.5 \cdot 10^{-9}$.

\subsection{SM meets $j_{S+P}J^{\dagger}_{S+P}$ and $j_{S+P}
J^{\dagger}_{S-P}$}
Under some assumptions \cite{paes97} one gets 
\be{me1}
ME_{S+P}^{S+P}=ME_{S-P}^{S+P}
=\frac{4}{R^2 m_e^2} {\cal M}_1^{(\nu)}
\ee
with the nuclear radius $R$
and $k=1$ determining the phase space factor. Inserting the numerical value
of the matrix element 
${\cal M}_1^{(\nu)}=2.1$, which
has been calculated in the QRPA approach in \cite{Hir96}, one derives 
$\epsilon^{S+P}_{S+P}=\epsilon^{S+P}_{S-P}<1.2 \cdot 10^{-8}$.  

\subsection{SM meets $j_{T_{R}}J^{\dagger}_{T_{R}}$ and $j_{T_{R}}
J^{\dagger}_{T_{L}}$}
In the tensor part the dacay rate depends on the phase space $k=1$ and
new matrix elements not considered
in the literature \cite{paes97}. An estimation of their numerical values for
the special case of $^{76}$Ge, based on kinematic properties, yields 
$\epsilon^{T_{R}}_{T_{R}}< {\cal O} (10^{-9})$. The limit on
$\epsilon^{T_{R}}_{T_{L}}$ is expected to be considerably weaker.

\section{Conclusion}
We have presented a general parametrization for the long range part of the
neutrinoless double beta decay rate in terms of effective couplings. 
Combined with the short range part and contributions of derivative 
couplings, this parametrization will give the double beta decay constraints
for arbitrary lepton number violating theories beyond the SM. 
A further step should include interference terms and justify the assumptions 
and approximations used in this work. For a more detailed discussion we 
refer to \cite{paes97}.

\section*{Acknowledgement}
H.P. is supported by a HSP/II AUFE grant of the Deutscher Akademischer 
Auslandsdienst (DAAD)

\end{document}